\documentclass[twocolumn]{aastex6}
\usepackage{graphicx,epstopdf}
\usepackage{enumerate}
\usepackage{natbib}

\newcommand{\Rosat}{\emph{ROSAT}}
\newcommand{\dxl}{\emph{DXL}}

\def\arcdeg{\hbox{$^\circ$}}

\received{}
\accepted{}
\shorttitle{The Structure of Local Hot Bubble}
\shortauthors{Liu, et al.}

\begin{document}

\title{The Structure of the Local Hot Bubble}
\author{W. Liu\altaffilmark{1}, M. Chiao\altaffilmark{2}, M. R. Collier\altaffilmark{2}, T. Cravens\altaffilmark{3}, M. Galeazzi\altaffilmark{1}, D. Koutroumpa\altaffilmark{4}, K. D. Kuntz\altaffilmark{5}, R. Lallement\altaffilmark{6}, S. T. Lepri\altaffilmark{7}, D. McCammon\altaffilmark{8}, K. Morgan\altaffilmark{8}, F. S. Porter\altaffilmark{2}, S. L. Snowden\altaffilmark{2}, N. E. Thomas\altaffilmark{2}, Y. Uprety\altaffilmark{1,9}, E. Ursino\altaffilmark{1,10}, B. M. Walsh\altaffilmark{11}}

\altaffiltext{1}{Department of Physics, University of Miami, Coral Gables, FL, 33124, U.S.A.}
\altaffiltext{2}{NASA Goddard Space Flight Center, Greenbelt, MD,  20771, U.S.A.}
\altaffiltext{3}{Department of Physics and Astronomy, University of Kansas, Lawrence, KS 66045, U.S.A.}
\altaffiltext{4}{Universite Versailles St-Quentin; Sorbonne Universites, UPMC Univ. Paris 06 \& CNRS/INSU, LATMOS-IPSL, 78280, France}
\altaffiltext{5}{The Henry A. Rowland Department of Physics and Astronomy, Johns Hopkins University, Baltimore, MD 21218, U.S.A.}
\altaffiltext{6}{GEPI Observatoire de Paris, CNRS, Universite Paris Diderot, 92190, Meudon, France}
\altaffiltext{7}{Department of Atmospheric, Oceanic, and Space Sciences, University of Michigan, Ann Arbor, MI 48109, U.S.A.}
\altaffiltext{8}{Department of Physics, University of Wisconsin, Madison, WI 53706, U.S.A.}
\altaffiltext{9}{Current address: Department of Physics and Astronomy, Middle Tennessee State University, Murfreesboro, TN 37132}
\altaffiltext{10}{Current address: Physics Department, Grinnell College, Grinnell, IA 50112}
\altaffiltext{11}{Department of Mechanical Engineering, Boston University, Boston, MA 02215, U.S.A.}
\email{galeazzi@physics.miami.edu}
\begin{abstract}
\dxl\ (Diffuse X-rays from the Local Galaxy) is a sounding rocket mission designed to quantify and characterize
the contribution of Solar Wind Charge eXchange (SWCX) to the Diffuse X-ray Background and study the properties of
the Local Hot Bubble (LHB). Based on the results from the \dxl\ mission,
we quantified and removed the contribution of SWCX to the diffuse X-ray background measured by the \Rosat\ All Sky Survey (RASS). The ``cleaned'' maps were used to investigate the physical properties of the LHB. Assuming thermal ionization equilibrium, we measured a highly uniform temperature distributed around $kT$=0.097 keV$\pm$0.013 keV (FWHM)$\pm$0.006 keV (systematic). We also generated a thermal emission measure map and used it to characterize the three-dimensional (3D) structure of the LHB which we found to be in good agreement with the structure of the local cavity measured from dust and gas.
\end{abstract}
\keywords{ISM: bubbles - ISM: structure - X-rays: diffuse background - X-rays:
ISM}

\section{Introduction}
The diffuse soft X-ray background observed at 1/4 keV in the \Rosat\ R12 band \citep{Snowden97} is dominated by a local source that shows no sign of absorption by cool interstellar gas
 \citep{Juda91}. One optical depth at 1/4 keV is roughly 1$\times$10$^{20}$ HI cm$^{-2}$, a quantity reached within ~50 pc at average interstellar densities. An irregular ``local cavity'' extending about 100~pc from the Sun was shown by the Copernicus satellite to be almost entirely devoid of cool gas \citep{Savage72,Jenkins74,Knapp75}. If filled with 10$^6$~K gas at a reasonable pressure, the cavity could produce observed ``local'' X-rays \citep{Sanders77}. The portion of the local cavity filled with this hot gas was dubbed the Local Hot Bubble (LHB) \citep{Sanders77,Tanaka77,Cox86}, and the enhanced X-ray emitting areas at intermediate latitudes were found to correlate well with minima in the measured neutral gas column \citep{Snowden90}, as if the cool gas had been displaced by the hot gas. \Rosat\ demonstrated that a smaller portion of the soft X-ray background is due to the Galactic halo
 \citep{Burrows91,Snowden91}. Emission from the hot Galactic halo contributes significantly only in areas of low absorption at intermediate and high Galactic latitudes.

This simple picture was upset by the discovery of diffuse X-ray emission from within the solar system due to Solar Wind Charge eXchange (SWCX), which could provide some or all the soft diffuse X-ray emission at 1/4 keV \citep{Cravens00,Cravens01,Robertson03,Lallement04,Koutroumpa09a}. SWCX emission is generated when the highly charged solar wind ions interact with the neutral materials within the solar system, gaining an electron in a highly excited state which then decays emitting an X-ray photon with the characteristic energy of the ion. In order to improve our understanding of the local diffuse X-ray emission and the structure of the LHB, it is essential to remove the contamination of the SWCX. However, despite many efforts, an accurate estimation of the SWCX is quite difficult, especially in the 1/4 keV band, due to the poorly known cross sections for producing the many X-ray lines from SWCX, limited data on heavy ion fluxes in the Solar Wind, and the general spectral similarity of SWCX and thermal emission \citep{Cravens00, Lallement04, Koutroumpa06, Snowden09, Henley08, Crowder12, Yoshino09}. Efforts to estimate the SWCX contribution to historical measurements for the diffuse X-ray background, such as in the \Rosat\ All-Sky Survey (RASS) are even more problematic due to the limited solar wind data \citep[see][for further discussion]{Kuntz15}.

\dxl\ \citep{Galeazzi11, Galeazzi12, Thomas13} is a sounding rocket mission designed to quantify and characterize the contribution of SWCX to the diffuse X-ray emission. To separate the SWCX contribution, \dxl\ uses the spatial signature of SWCX emission due to the ``helium focusing cone'', a higher neutral He density region downwind of the Sun \citep{Michels02, Snowden09, Thomas13}. By comparing the \dxl\ data and the RASS data along the \dxl\ scan path, our team measured the broad band averaged cross sections and provided a significantly more accurate empirical estimate of the SWCX emission. \dxl\ estimated the total SWCX contribution to be $\sim40$\% of the X-ray flux at 1/4 keV in the Galactic plane \citep{Galeazzi14}, supporting the previous picture of a hot bubble filling the local interstellar medium in all directions and accounting for the remaining $\sim$60\% in the plane. Based on the results from \citet{Galeazzi14}, \citet{Snowden14} showed that the gas pressure from the remaining local emission is in pressure equilibrium with the local interstellar clouds, eliminating the long standing pressure problem of the LHB \citep{Jenkins09}.

In this paper we reevaluate the properties of the LHB based on the RASS data \citep{Snowden97} combined with the estimate of the SWCX contribution from \dxl. We focused on the R1 ($\sim$0.11-0.284 keV) and R2 ($\sim$0.14-0.284 keV) data, as the LHB contribution to the R4 ($\sim$0.44-1.01 keV) and R5 ($\sim$0.56-1.21 keV) bands is negligible. In \S~2 we describe how to remove the SWCX emission from the RASS data, and to estimate the LHB temperature and emission measure, \S~3 contains the results, and conclusions are in \S~4.

\section{Data Analysis}
\label{dataanalysis}

\citet{Snowden98, Snowden00} used the shadows cast by nearby (100-200~pc) clouds to estimate and remove the contribution from background emission (Galactic halo and extragalactic components) to the RASS R1 and R2 maps, producing
clean ``local'' maps ($<$100-200~pc). With the advance in X-ray telescopes, the shadow technique is now feasible for individual pointing for spectroscopy study to disentangle the foreground and background \citep{Galeazzi07,Smith07,Gupta09b,Henley15,Liu16,Ursino16}. These ``local'' maps should contain only the contribution from SWCX, both heliospheric ($S(\ell,b,t)$) and geocoronal ($G$), and the LHB ($L(\ell,b)$). For each RASS band we can therefore write the total flux, $F(\ell,b)$, as:
\begin{equation}
F(\ell,b,t) = S(\ell,b,t) + L(\ell,b)+G
\end{equation}

Following the procedure from \citet{Uprety16}, based on the models of \citet{Koutroumpa06}, the heliospheric component can be written as
\begin{equation}
S(\ell,b,t)=\beta(t) \times N(\ell,b)
\end{equation}
where
\begin{equation}
\beta(t)=n_p(R_0,t)v_{rel}(t)\alpha_{He}
\end{equation}
depends on the solar wind properties and the cross section with neutrals ($n_{p}(R_{0},t)$ is the proton density at $R_{0}=1AU$, $v_{rel}$ is the relative speed between solar wind and neutral flow, and $\alpha_{He}$ is the compound cross-section for Helium), and
\begin{equation}
N(\ell,b)=\int\frac{n_{He}}{R^2}ds+\frac{\alpha_H}{\alpha_{He}}\int\frac{n_H}{R^2}ds
\end{equation}
where $\int\frac{n}{R^2}ds$ is the integrated neutral column density along the line of sight, weighted by one over the distance from the Sun squared, and $\frac{\alpha_H}{\alpha_{He}}$ is the ratio between cross sections with H and He.

\begin{figure*}
\epsscale{1.15}
\plottwo{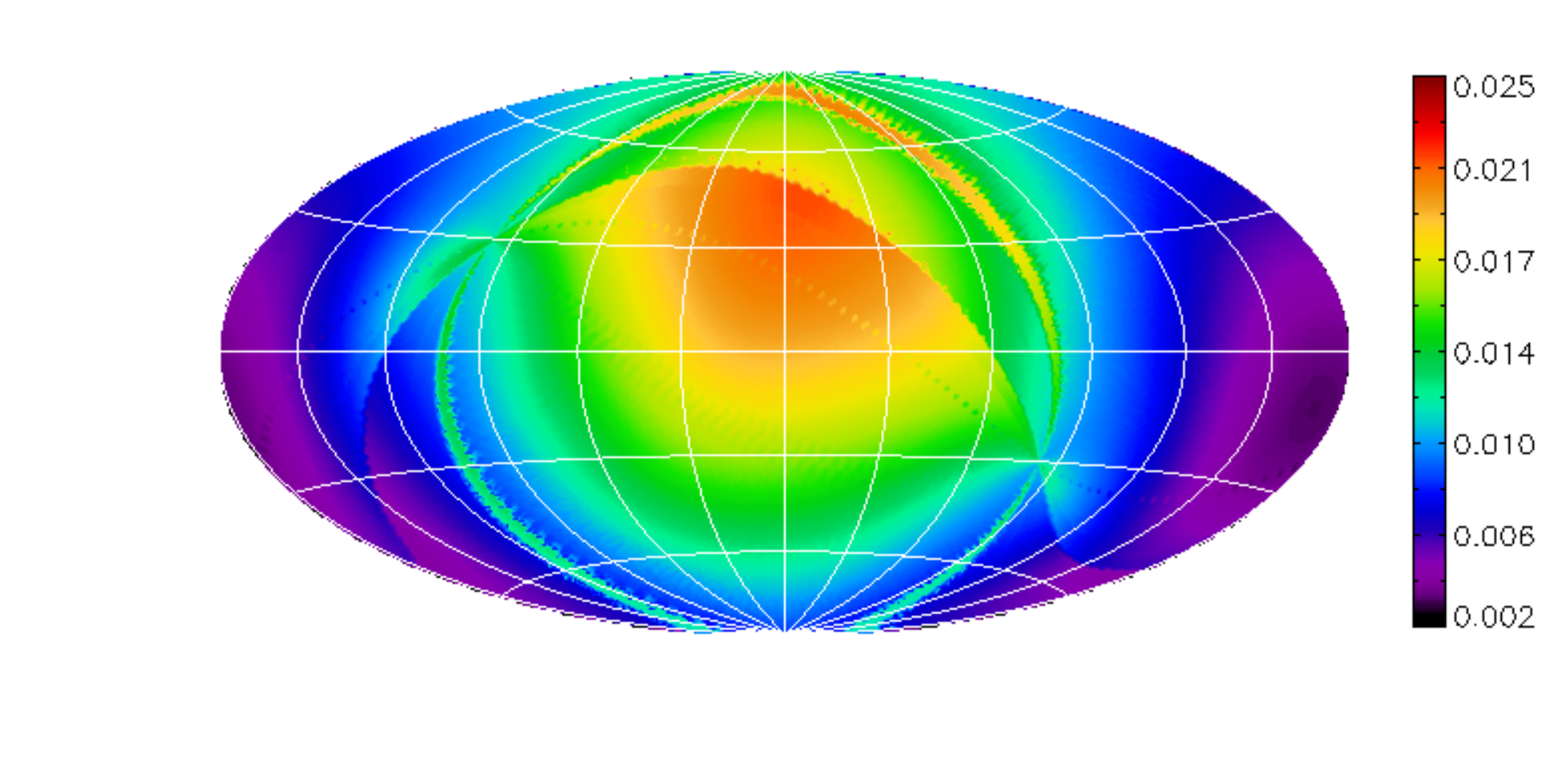}{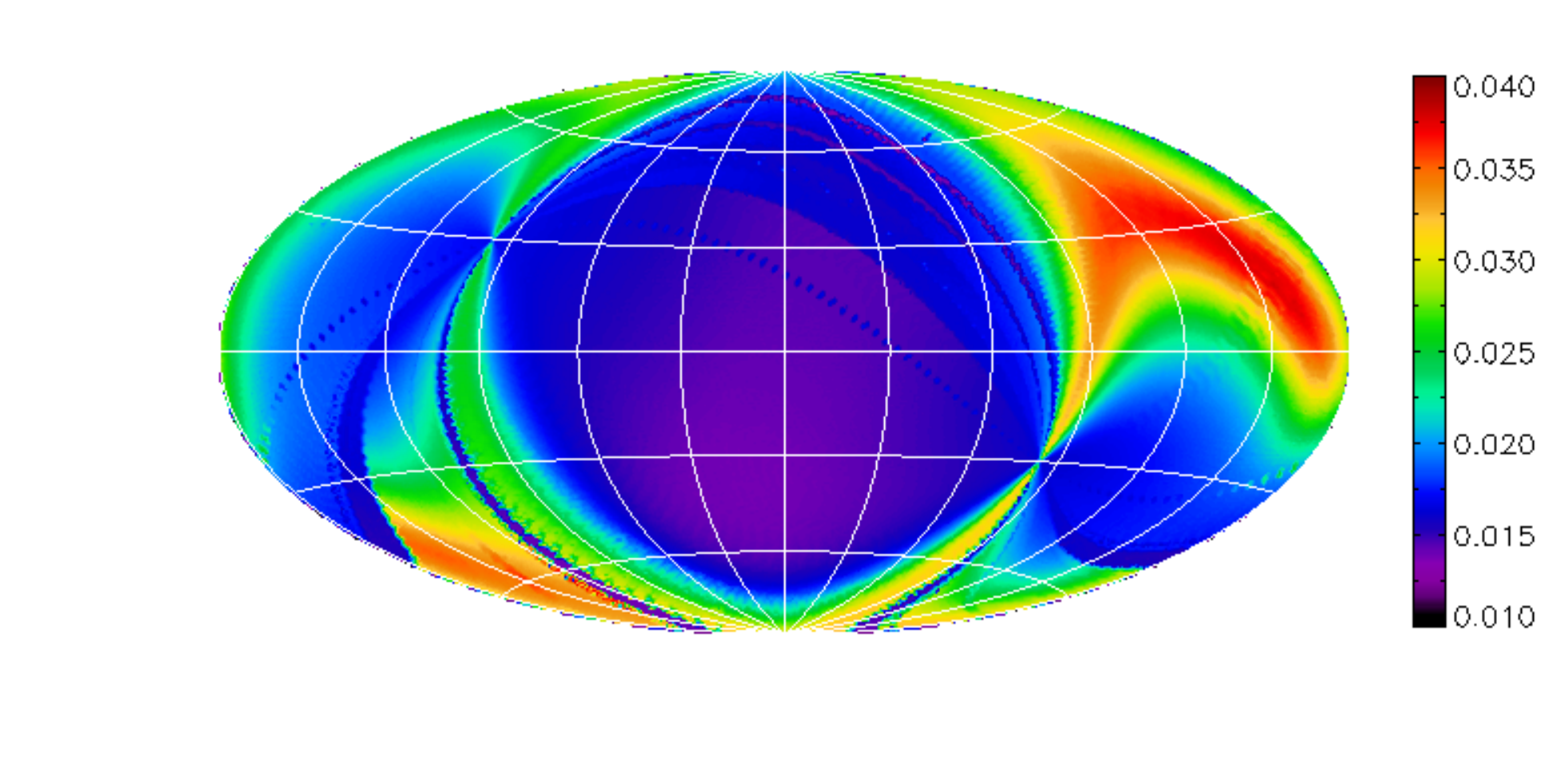}
\caption{The Aitoff-Hammer projection of the neutral integral distribution for H (left) and He (right) in units of cm$^{-3}$AU$^{-1}$.
\label{neutral}}
\end{figure*}

\citet{Uprety16} combined data from the same part of the sky from \dxl\ and RASS and found the best fit parameter for $\beta(t)$ for each RASS band for given values of $G$ and $\alpha_H/\alpha_{He}$. Therefore, the heliospheric SWCX contribution to any RASS band for any direction can be directly estimated given the neutral distribution.
Figure~\ref{neutral} shows the Aitoff-Hammer projection of the H and He neutral integral during the RASS campaign, calculated based on a well determined model for the interstellar neutral distributions within the solar system \citep{Lallement85a,Lallement85b,Lallement04,Koutroumpa06}. The sharp edges visible in this maps are due to abrupt shifts in vantage point around the Earth's orbit during the \Rosat\ survey, since the survey comes back to its starting point after six months and there were missed sections that were backfilled at later times.

\begin{figure*}
\epsscale{1}
\plotone{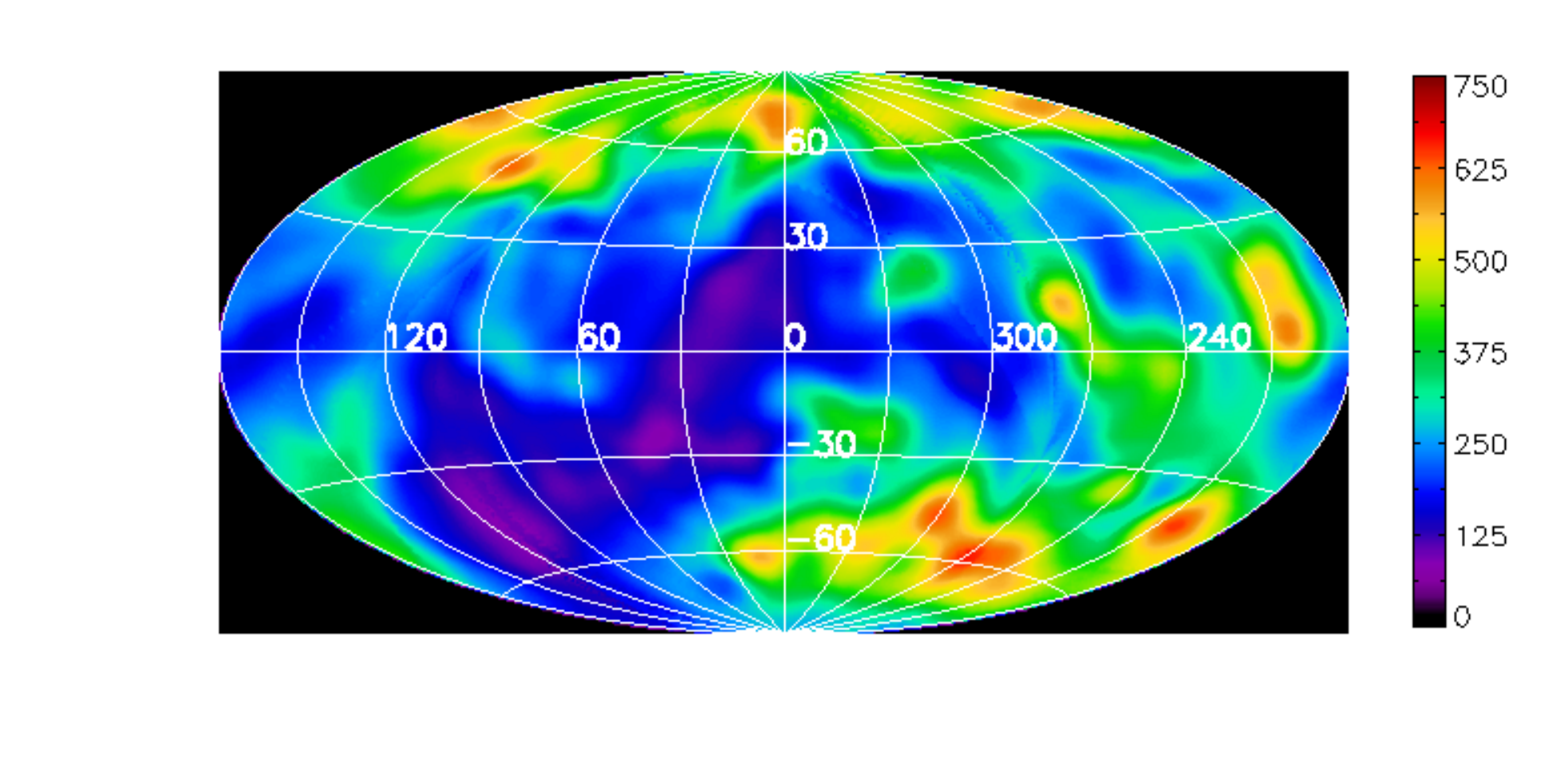}
\caption{The Aitoff-Hammer projection of the total ``cleaned'' LHB emission in the RASS R1+R2 band in RU after removing both the non-local components and SWCX contribution.
\label{R1plusR2}}
\end{figure*}

\begin{figure*}
\epsscale{1}
\plotone{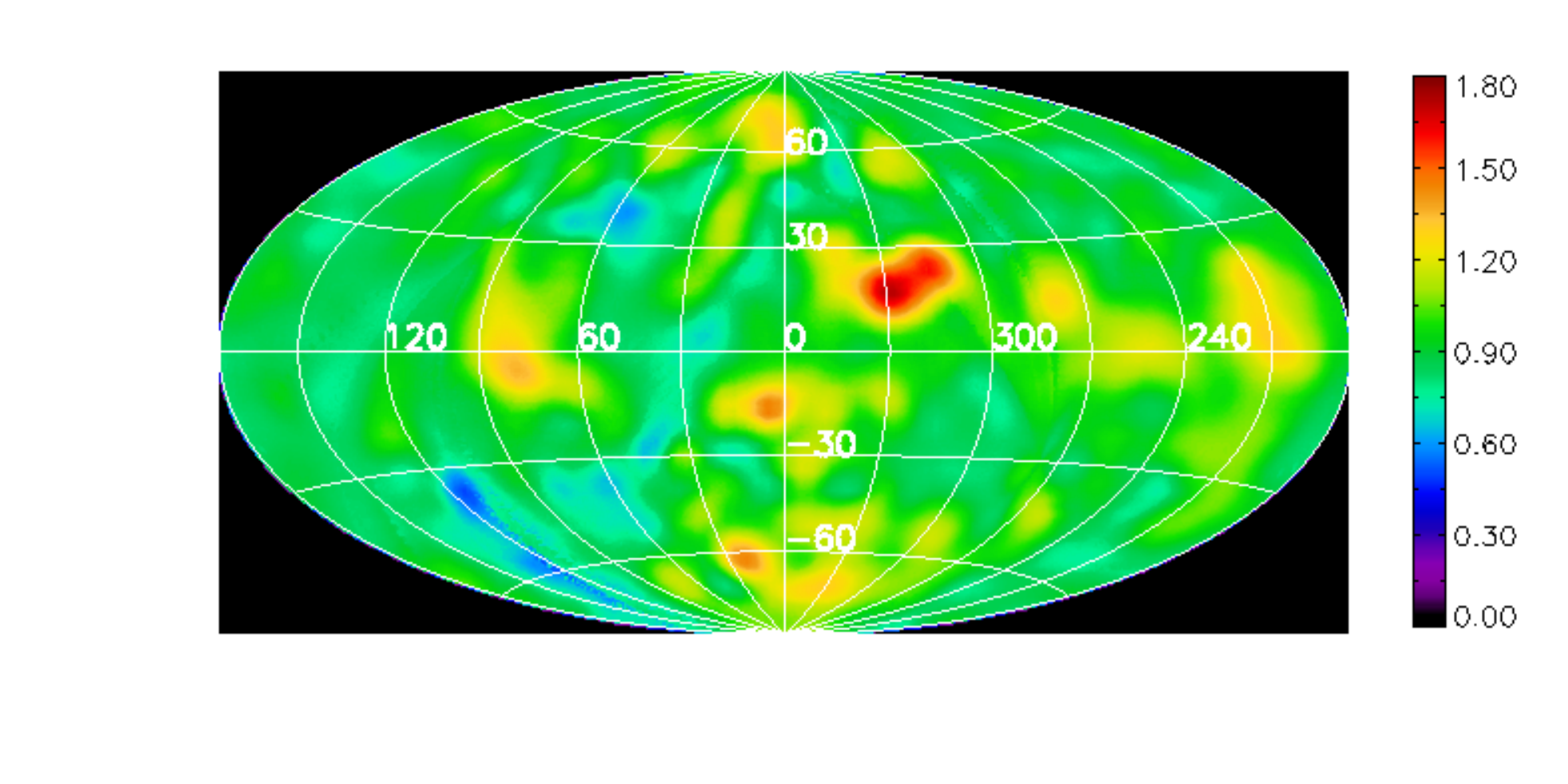}
\caption{The Aitoff-Hammer projection of the R2/R1 band ratio of the ``cleaned'' LHB.
\label{R2R1ratio}}
\end{figure*}

\citet{Uprety16} extensively discuss the use of different combinations of $G$ and $\alpha_H/\alpha_{He}$ and their effect of the systematic error of their results. For this investigation, we use their best fit parameters for $\beta(t)$ shown in their Table~2, combined with the neutral distribution shown in Figure \ref{neutral} to generate maps of SWCX contribution to both R1 and R2. The model SWCX maps were then subtracted from the local maps of \citet{Snowden98, Snowden00} to produce the ``clean'' LHB maps in the R1 and R2 band which contain only the LHB emission. We note that these maps are different from those shown in \citet{Uprety16}. \citet{Uprety16} subtracted the SWCX contribution for the total R1 and R2 bands, producing ``clean'' astrophysical maps, containing both local and non-local components. Figure~\ref{R1plusR2} shows the Aitoff-Hammer projection of the total LHB emission in R1+R2 band, and Figure~\ref{R2R1ratio} shows the projection of the R2/R1 ratio of the LHB (notice that all the maps are smoothed since the LHB emission is assumed to vary smoothly over large angular scales and our analysis is insensitive to any variation on finer scales).

\begin{figure}
\epsscale{1.2}
\plotone{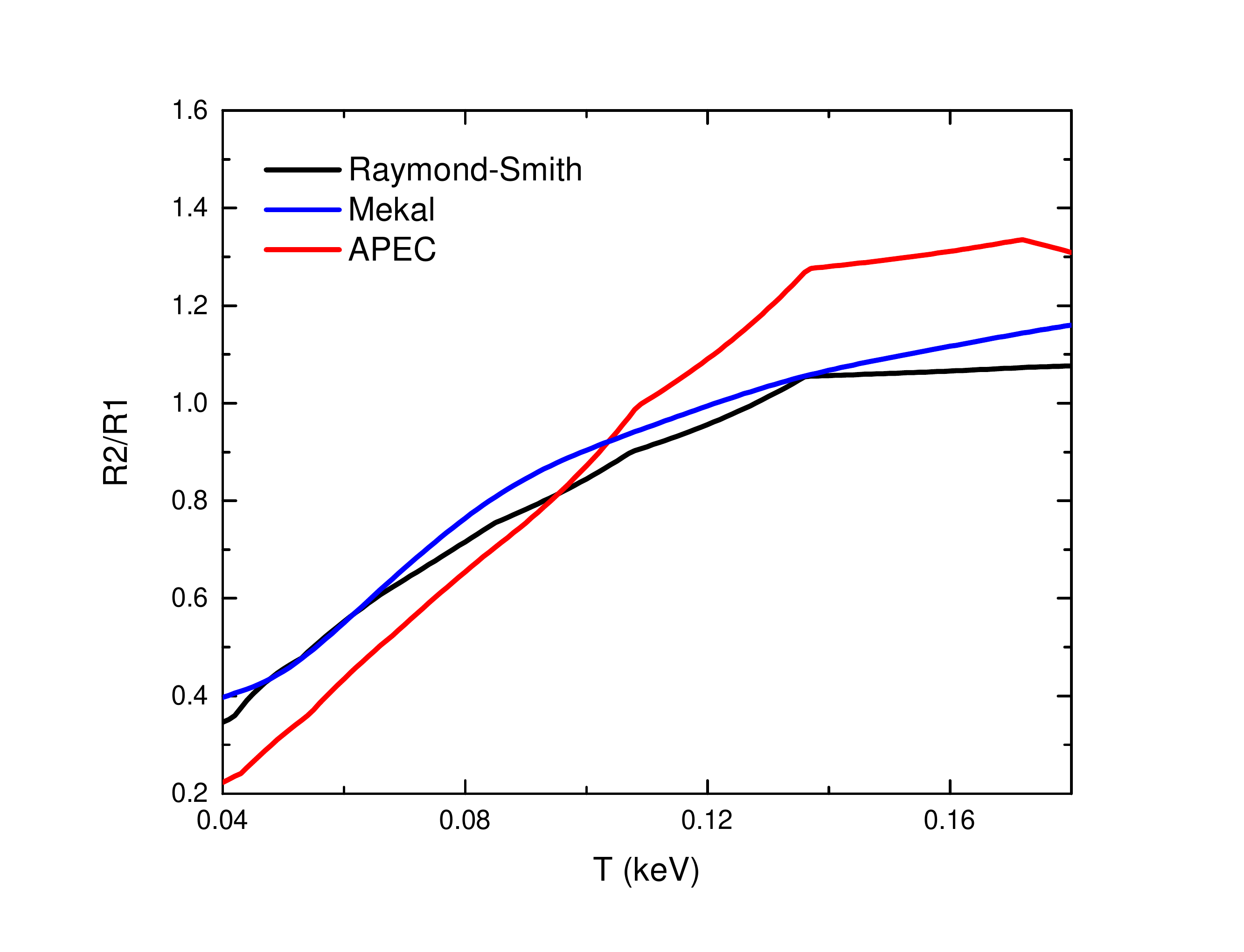}
\caption{R2/R1 ratio as a function of temperature from Raymond-Smith model (black), Mekal model (blue), and APEC model (red) with \citet{Anders89} abundance table.
\label{TR2R1}}
\end{figure}

Assuming that the LHB is in collisional equilibrium and can be well represented by a single, unabsorbed thermal component, it is possible to estimate the temperature of the LHB in any given direction based on the ratio of the R2/R1 bands. Unfortunately, none of the current thermal models available in XSPEC \footnote{http://heasarc.gsfc.nasa.gov/xanadu/xspec/} are particularly accurate in the R1 and R2 bands.
The Raymond-Smith model \citep{Raymond77} estimates the emission of a large number of weak lines that are known to be present, but it lacks accurate excitation rates and wavelengths. The Mekal model \citep{Mewe85,Kaastra93} is identical in treatment of ionization balance with Raymond-Smith model, but has many more lines and updated Fe L calculations.
The APEC model \citep{apec} includes only transitions for which accurate atomic rates are available and lacks many lines at low energy. In Figure \ref{TR2R1} we plot the R2/R1 ratio as a function of temperature using the Raymond-Smith, Mekal, and APEC models with \citet{Anders89} abundance table. It is immediately evident that the curve for APEC model is quite different from the other two and therefore any conclusion will depend on the model used. However, as it turns out, the systematic effect introduced by the choice of model is not large as our LHB data, as we will show in the next section, are clustered in the region where the curves nearly overlap.

We also point out that in the RASS maps there are brighter regions associated with additional X-ray emission from extended sources, e.g., the Cygnus Loop, Vela SNRs, the Galactic halo beyond the Draco Clouds, and the Monogem Ring. We have excluded them in our study by both setting an upper limit in the RASS R4+R5 value and manually removing regions associated with known structures unrelated to the LHB.

\begin{figure}
\epsscale{1.2}
\plotone{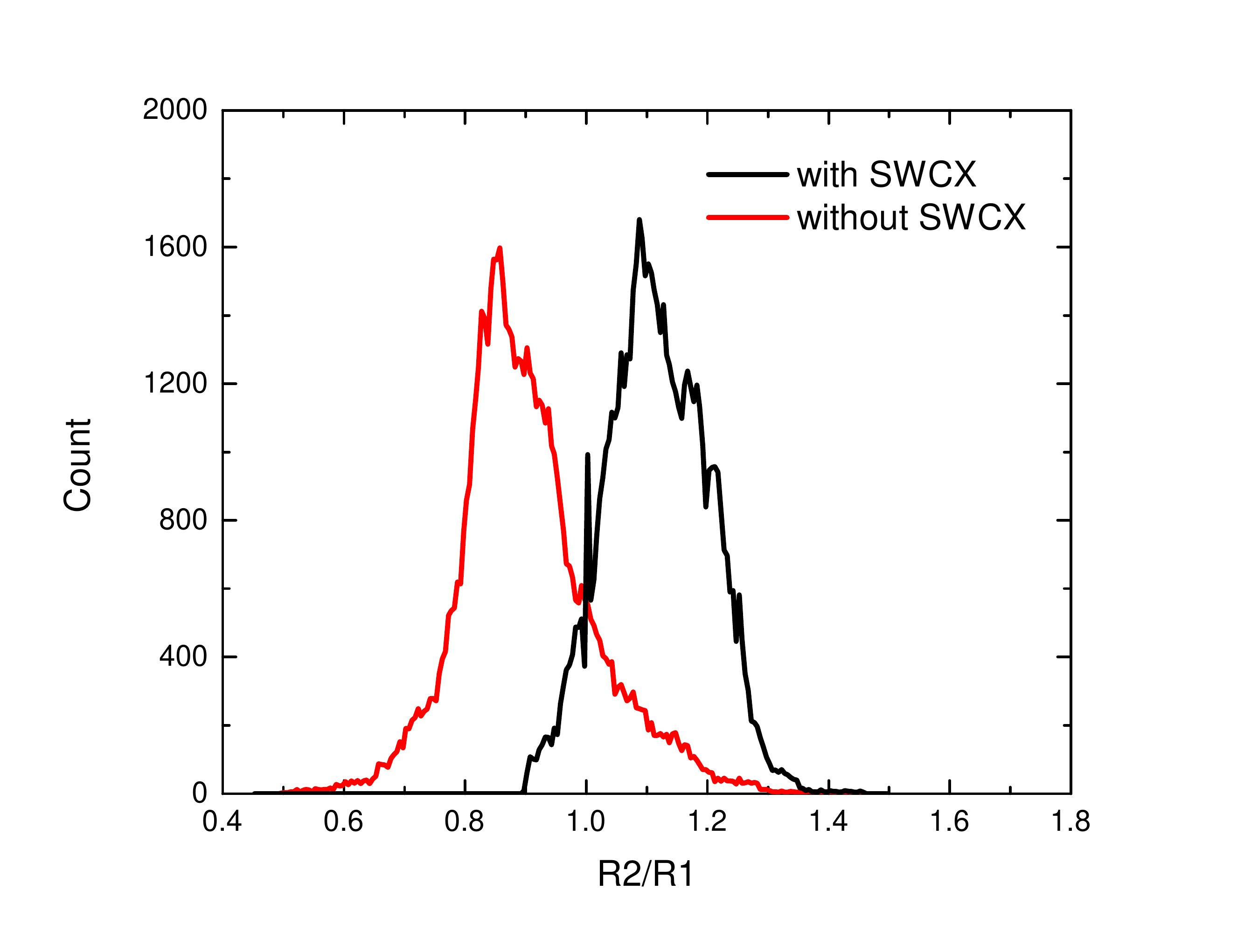}
\caption{The distribution of R2/R1 ratio before subtracting the SWCX (in black) and after subtracting the SWCX (in red).
\label{r2r1hist}}
\end{figure}

\section{Results}
\label{results}
\subsection{The LHB Temperature}
\begin{figure*}
\epsscale{1.16}
\plottwo{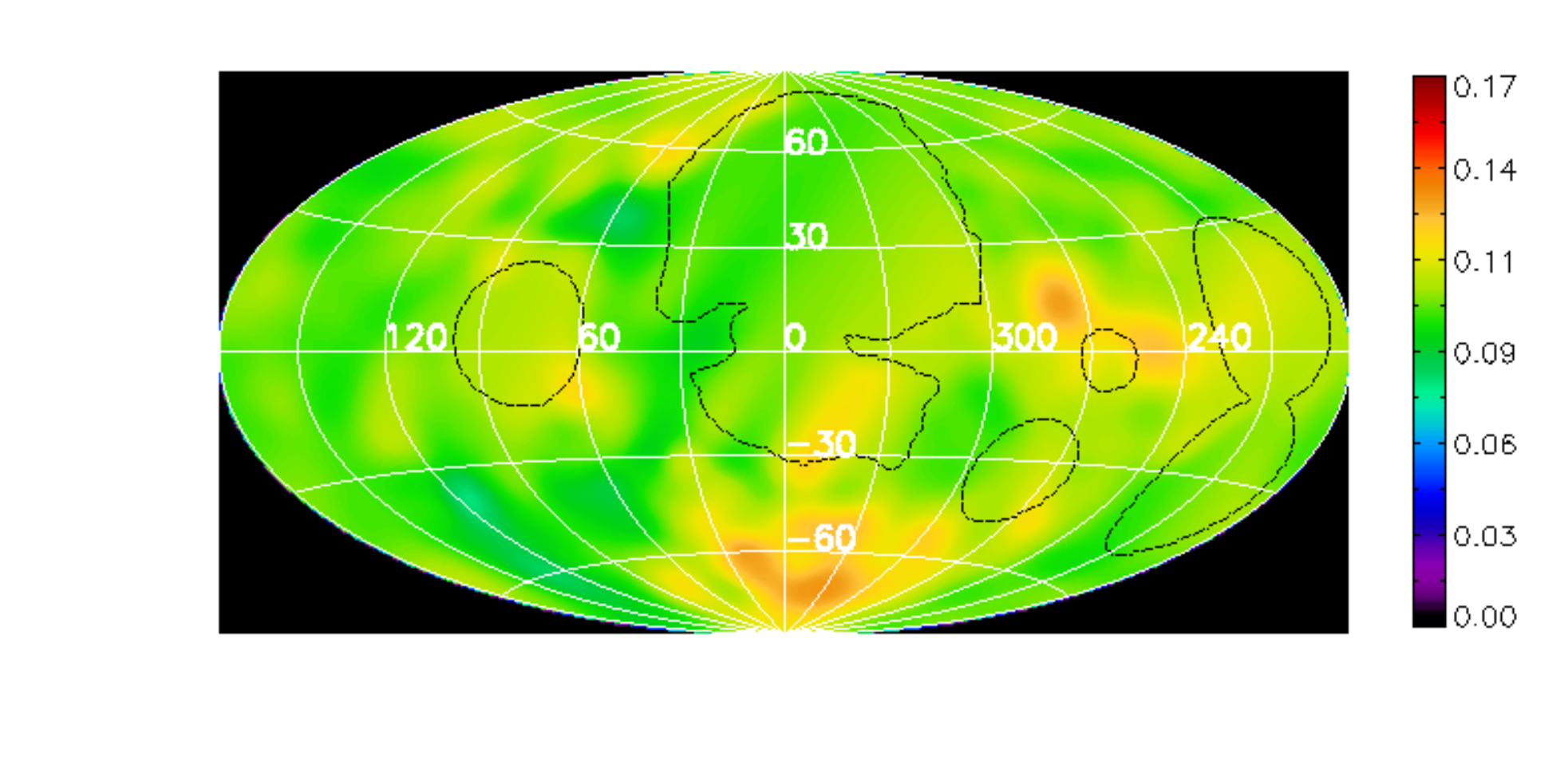}{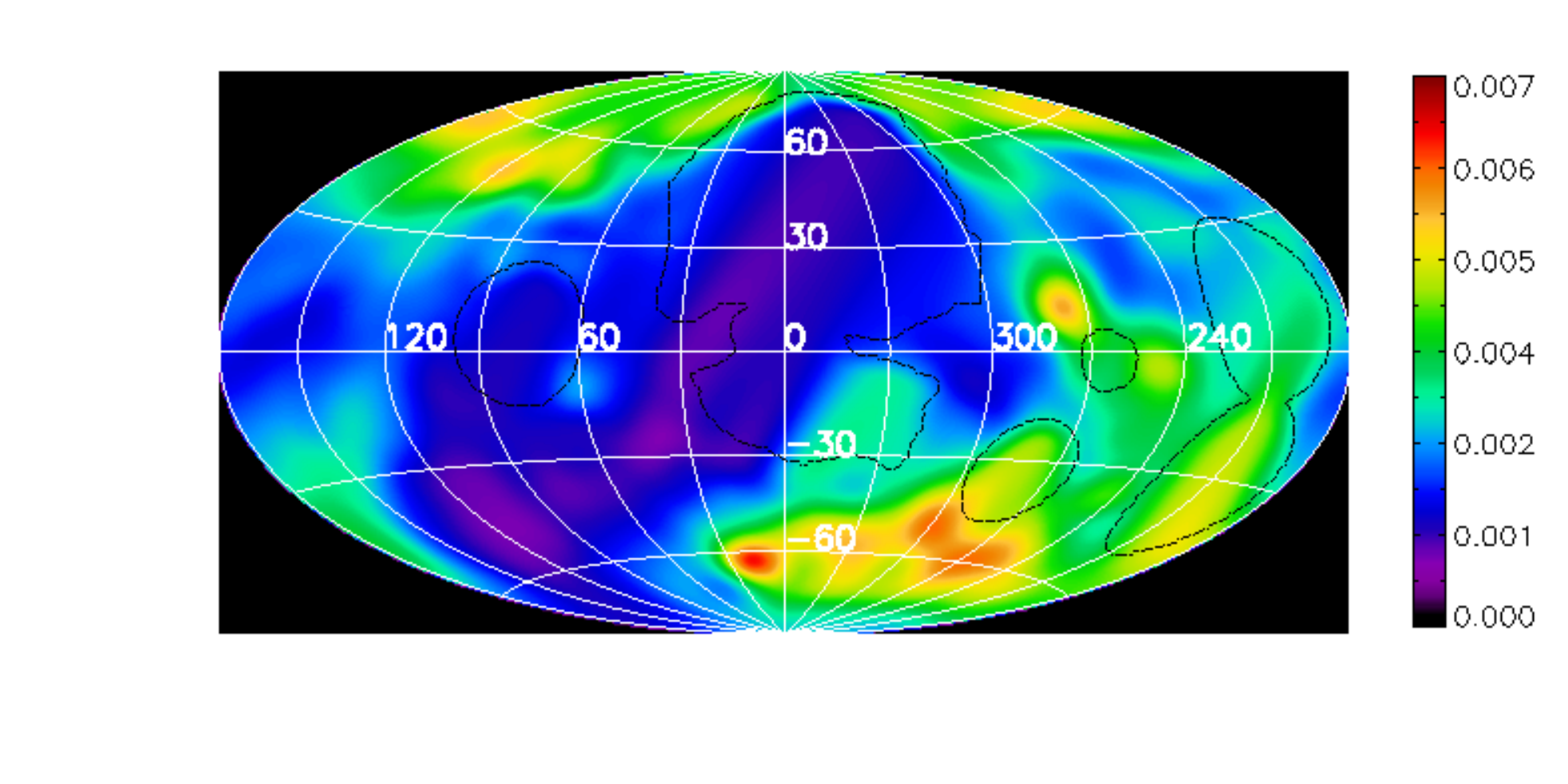}
\caption{The Aitoff-Hammer projection of the LHB temperature (left) in keV and the LHB emission measure in cm$^{-6}$ pc. Regions affected by non-LHB extended sources have been interpolated and circled in black.
\label{lhb}}
\end{figure*}
In Figure~\ref{r2r1hist} we show the distribution of the R2/R1 ratio over the whole sky before (in black) and after (in red) removing the SWCX contribution from the RASS data. The peak of R2/R1 is shifted from $\sim1.09$ to 0.86, with a FWHM of $0.16$. This corresponds to a temperature of $kT$=0.099 keV using the APEC model, 0.103 keV for Raymond-Smith model and 0.091 keV for Mekal model. The difference between APEC and Raymond-Smith models is very small since the R2/R1 ratio of 0.86 is in a region where the three lines in Figure~\ref{TR2R1} are very close to each other. Combining the three results, we estimate the peak temperature as $kT$=0.097 keV, with a systematic error of 0.006 keV. This systematic error also includes the systematic uncertainties in the SWCX parameters discussed in \citet{Uprety16}. Based on the the relation between the R2/R1, the temperature of the LHB is therefore cooler than previous estimates based on the maps without SWCX subtraction \citep{Snowden98,Kuntz00}.
In the left of Figure~\ref{lhb} we also show the Aitoff-Hammer projection of the temperature of the LHB. The distribution of the LHB temperature is quite uniform with a FWHM of 0.013 keV. In the figure, the areas affected by non-LHB extended sources have been interpolated and circled in black.

\subsection{The LHB Emission Measure and Size}

We also used the new R1 and R2 maps to extract the Emission Measure (EM) of the emitting plasma for each direction in the sky based on the APEC model. The Aitoff-Hammer projection of the LHB EM is shown on the right of Figure~\ref{lhb}. The emission measure is generally larger towards high latitude while smaller at low latitude in the northern hemisphere. In the southern hemisphere the emission measure is small from $0^{\circ}<$ l $< 180^{\circ}$. It is small at low latitude and large at high latitude from $180^{\circ}<$ l $< 360^{\circ}$. Over the whole sky, the distribution ranges from $\sim$0.8$\times$10$^{-3}$~cm$^{-6}$ pc to $\sim$6.5$\times$10$^{-3}$~cm$^{-6}$ pc.

\begin{figure*}
\epsscale{1}
\plottwo{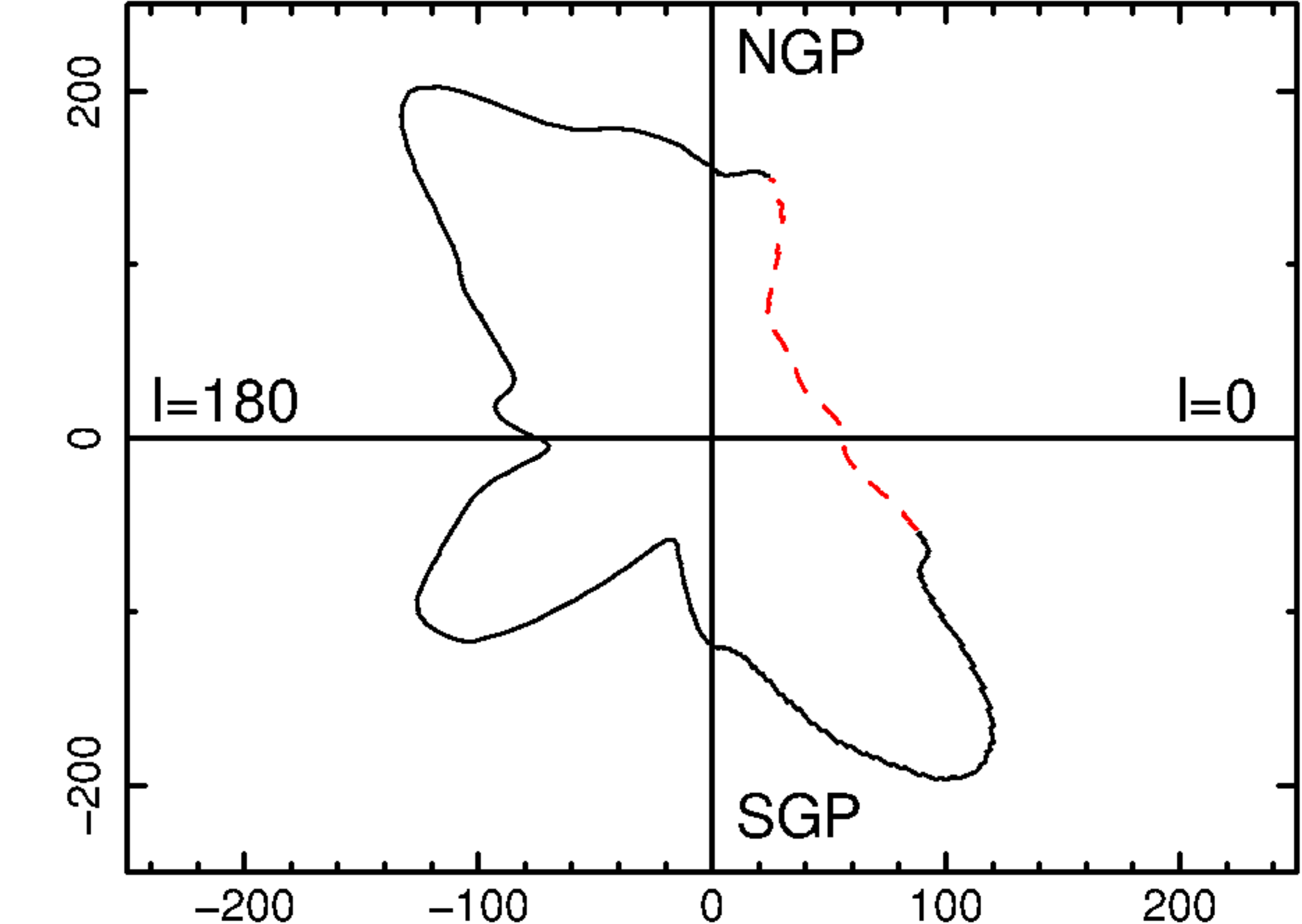}{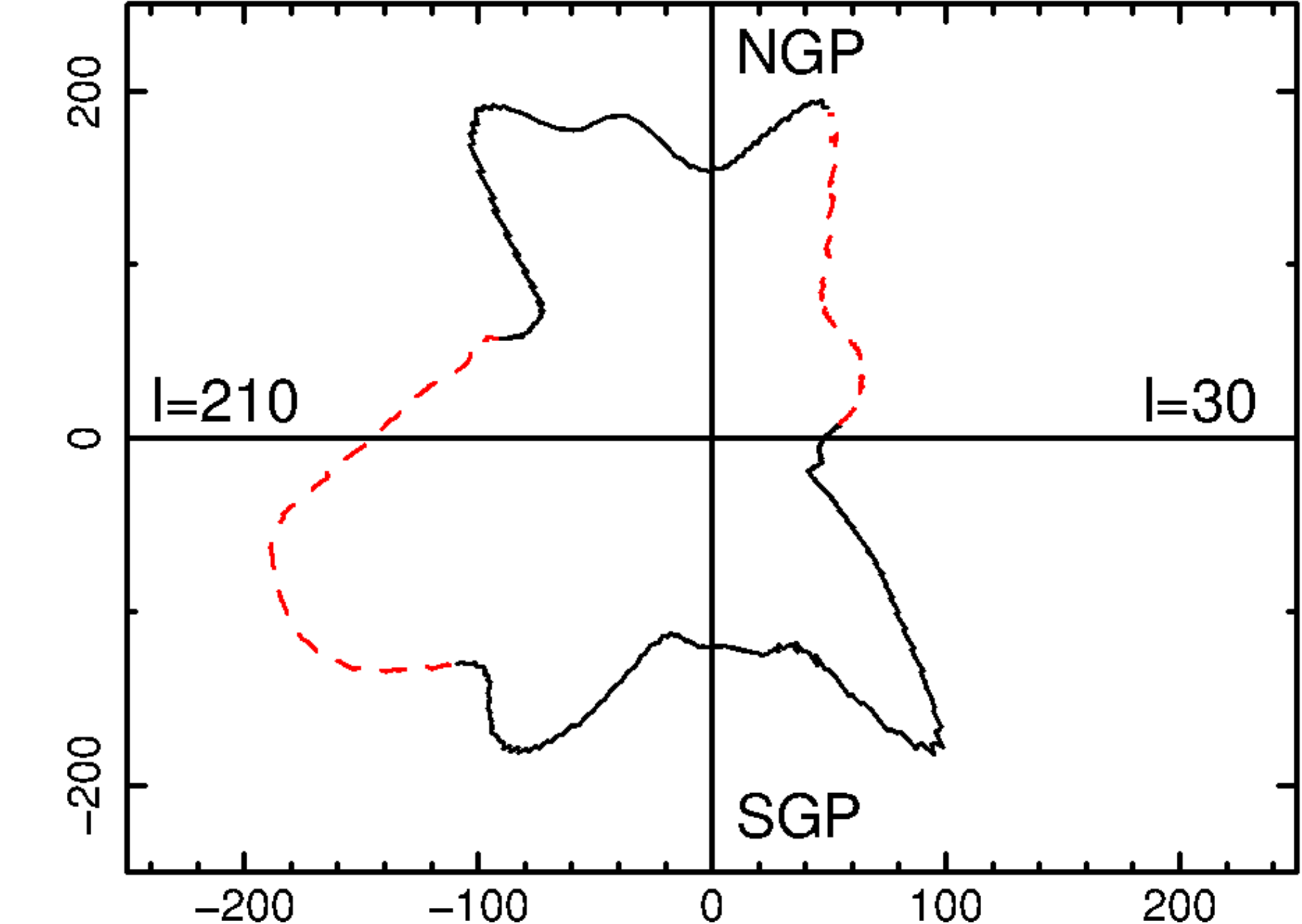}
\vspace{0.4cm}
\epsscale{1}
\plottwo{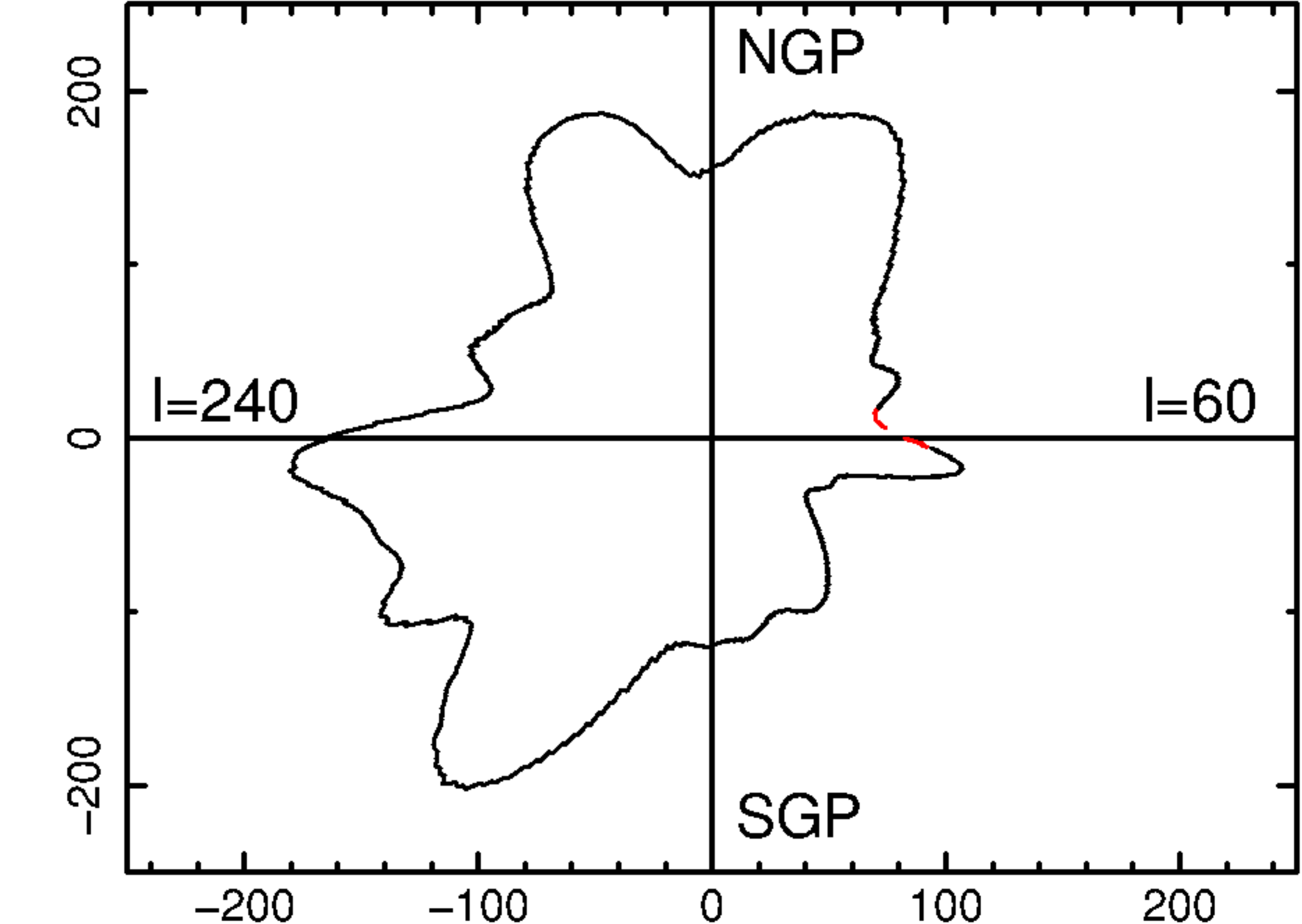}{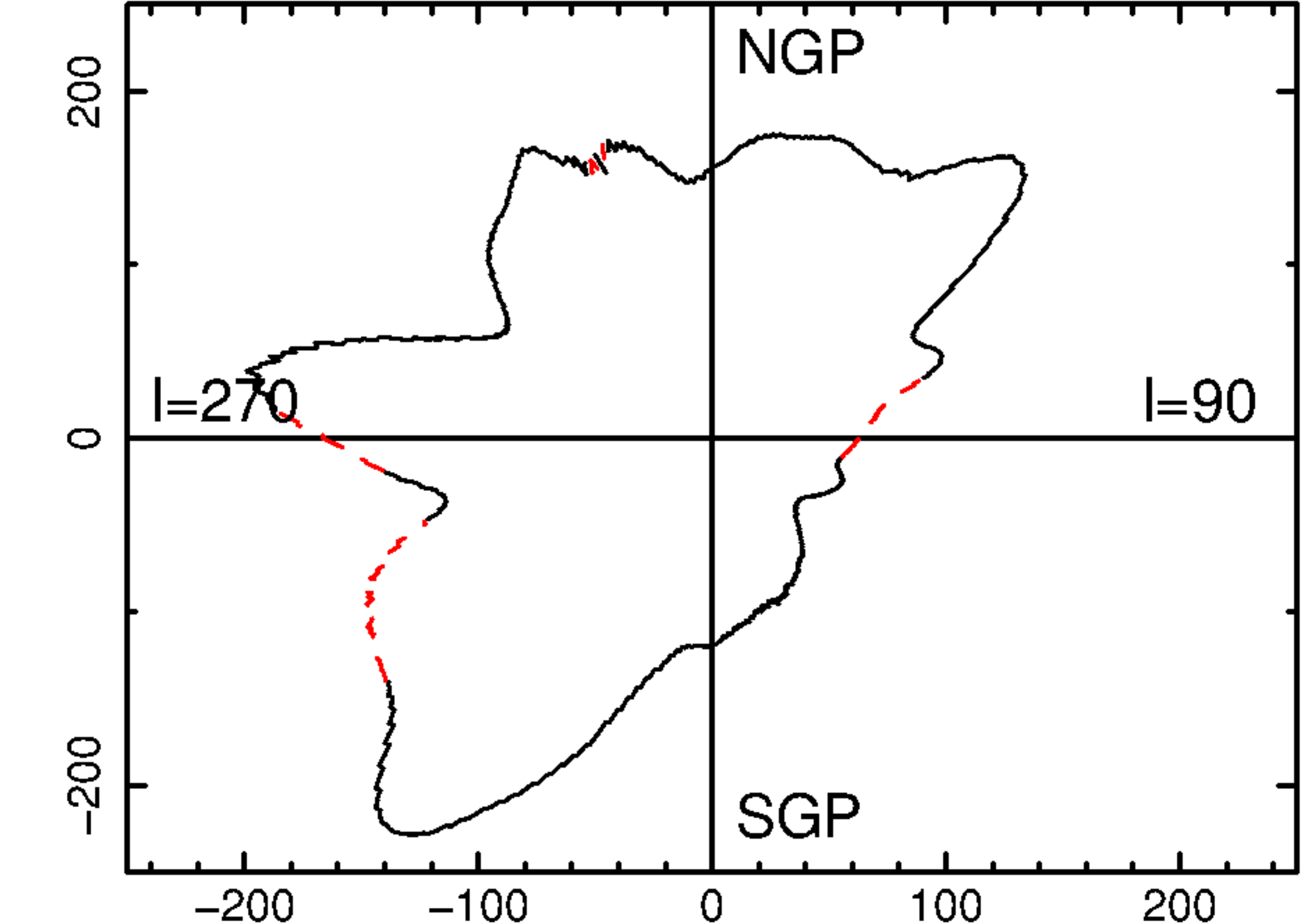}
\vspace{0.4cm}
\epsscale{1}
\plottwo{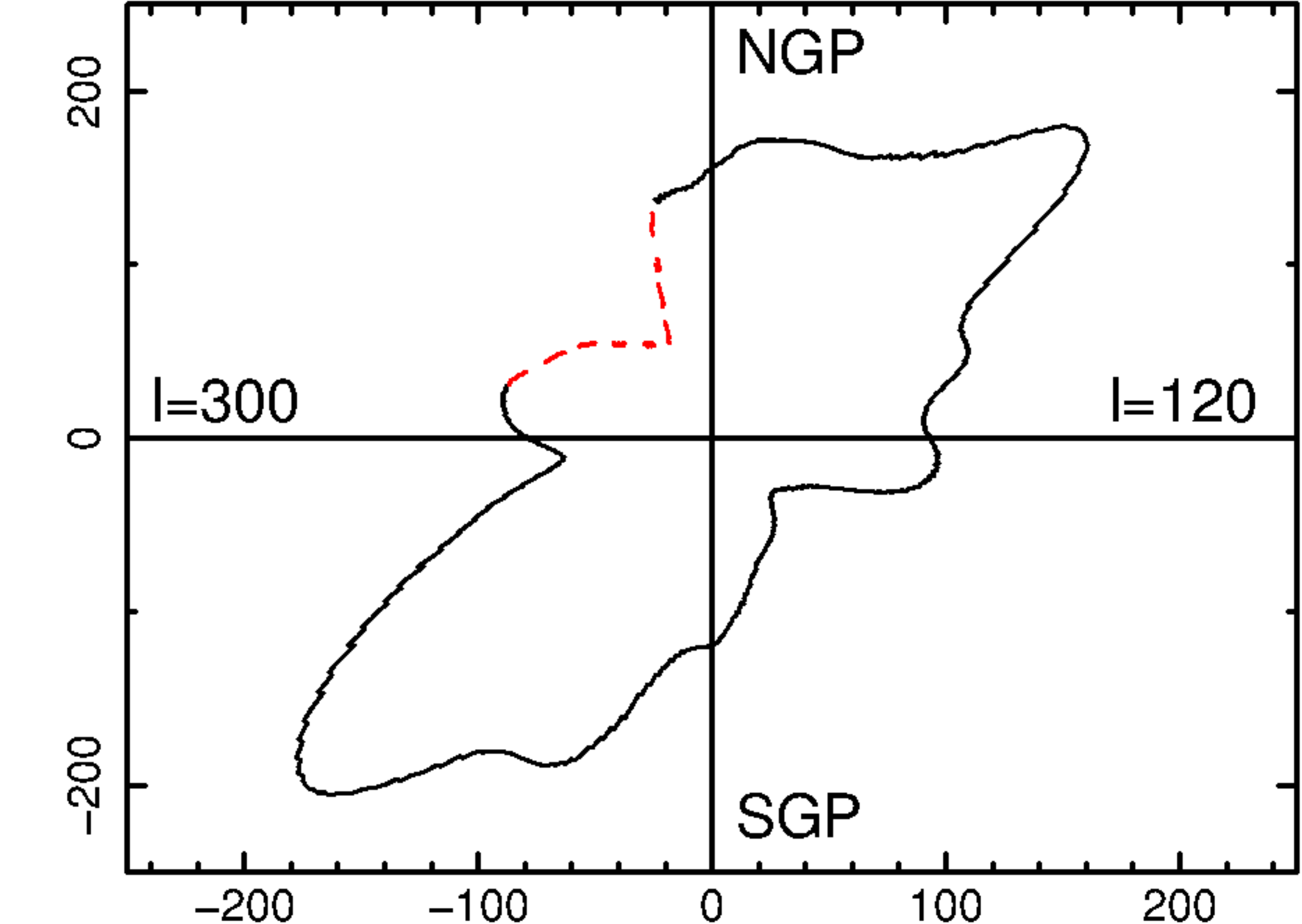}{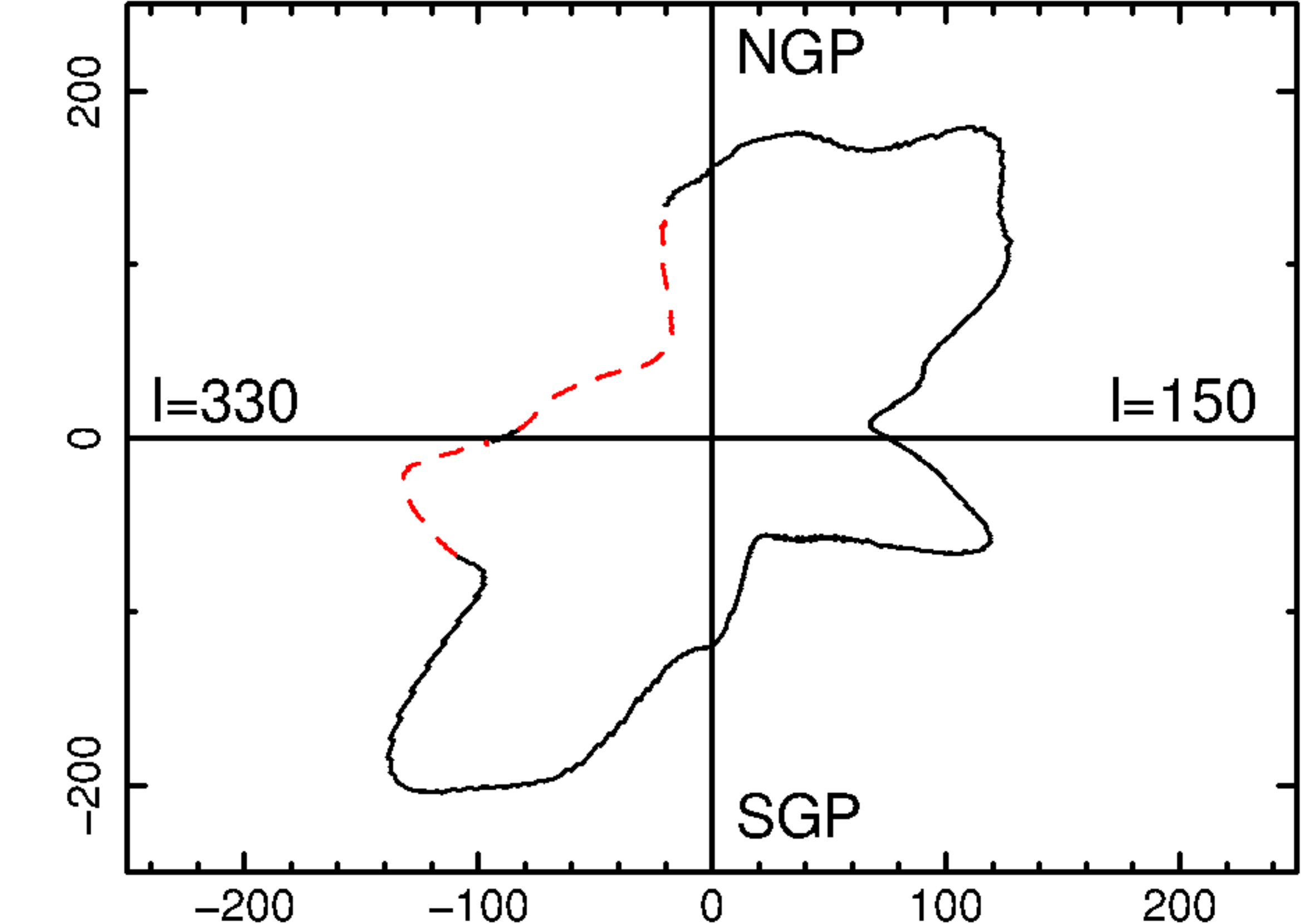}
\vspace{0.4cm}
\caption{The radius of the LHB in great-circle cuts through the Galactic poles along the labeled longitude. The
red dashed line corresponds to directions of non-LHB bright extended sources.
\label{lhb_pole}}
\end{figure*}

\begin{figure}
\epsscale{1}
\plotone{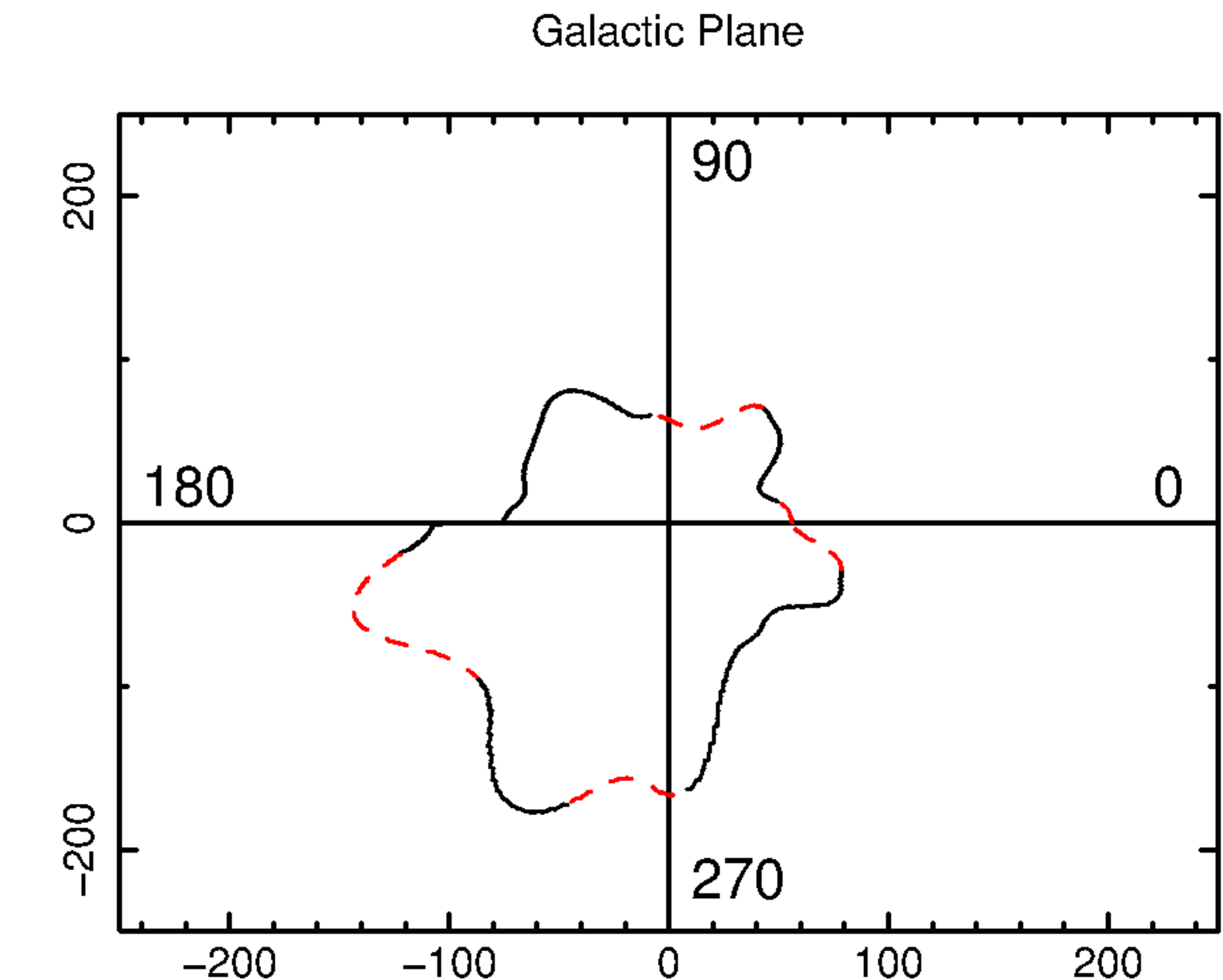}
\caption{The radius of the LHB in great-circle cuts through the Galactic plane. The
red dashed line corresponds to directions of non-LHB bright extended sources.
\label{lhb_plane}}
\end{figure}

Assuming that the electron density in the LHB is constant, we can use the EM to estimate the size of the LHB. For constant electron density, the emission measure is expressed as $EM=n_{e}n_{p}L$, where $n_{e}$ and $n_{p}$ are the electron and proton densities, $L$ is the path length through the LHB emitting plasma. Adopting the electron value of $n_{e}=4.68\times10^{-3}$ cm$^{-3}$ \citep{Snowden14}, we estimated the extension of the LHB in all directions and we built its three-dimensional structure. Figures \ref{lhb_pole} and \ref{lhb_plane} show the extension of the LHB along great-circle cuts through the Galactic pole and Galactic plane. The dash lines correspond to regions contaminated by distant bright sources \citep{Snowden98}.

\begin{figure}
\epsscale{0.9}
\plotone{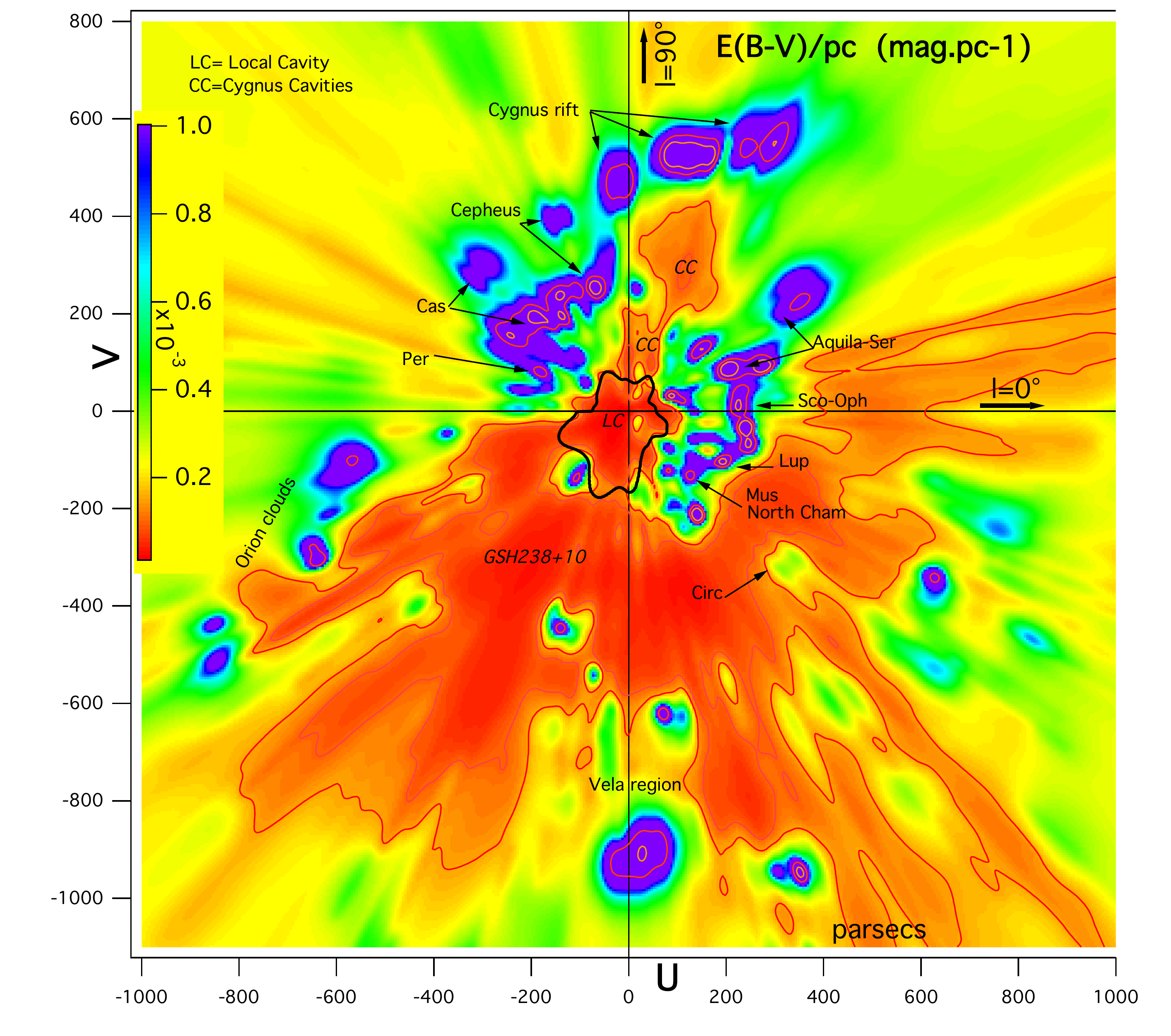}
\caption{Differential color excess shows the inverted differential opacity distribution in the Galactic plane (map is taken from \citet{Lallement14}). The black line shows the contour of the LHB in the Galactic plane from our measurements.
\label{lhb_contour_Galplane}}
\end{figure}

\begin{figure}
\epsscale{0.9}
\plotone{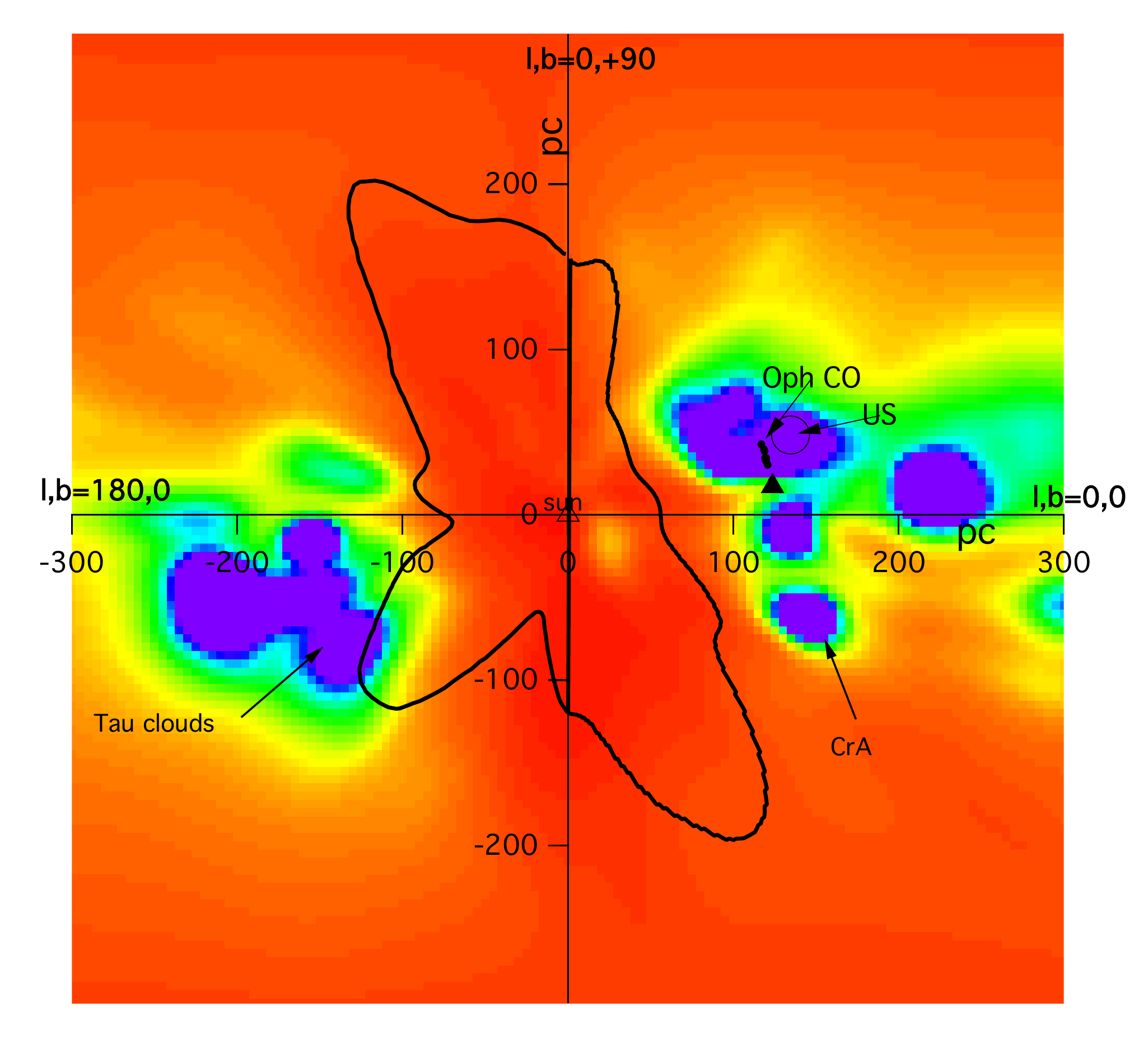}
\caption{Same as Figure \ref{lhb_contour_Galplane} but in the vertical plane and on a smaller scale.
\label{lhb_contour_vertical}}
\end{figure}

We also compared our results with measurements at other wavelenghts. Figure \ref{lhb_contour_Galplane} shows the distribution of the local Inter-Stellar Medium (ISM) in the Galactic plane from reddening data \citep{Lallement14}. The superimposed black line represents the contour of the LHB from our measurement which is the same as in Figure \ref{lhb_plane}. The shape of the LHB matches the boundary of the local cavity very well after removing the contribution of SWCX showing that the LHB and the local cavity are closely correlated. Figure \ref{lhb_contour_vertical} shows the same data as Figure \ref{lhb_contour_Galplane} but in the vertical plane and on a smaller scale. Although there is no clear boundary information of the local cavity toward high latitude, our contour matches well the local cavity at low latitude.

Based on our reconstruction of the LHB, we also calculated the total energy currently enclosed in the LHB as ~3.38$\times$10$^{50}$ ergs which is about 15.6 times smaller than estimated without removing the SWCX contribution to the RASS maps. We note that, while this is consistent with the energy released in a single supernova explosion, the LHB has been cooling away for millions of years, and its size and longevity remain inconsistent with a single SN explosion \citep{Cox82,Cox86,Smith01}.

There are a few systematics which affects our results. The first is the choice of model to convert the R1 to R2 ratio to temperature, which has already been discussed, and contributes a systematic error of $kT$=0.006 keV to the estimate of the LHB temperature. We also investigated the effect of using different abundance tables. For example, we used the abundance table by \citet{Wilms00} and foud it has a smaller effect than the choice of model, contributing a systematic error of $kT$=0.003~keV. The second major source of systematic uncertainty is the assumption made to derive the value of $\beta(t)$ by fitting the \dxl\ data, namely the choice of $\alpha_H/\alpha{He}$ and $G$. As discussed in \citet{Uprety16}, however, although the value of $\beta(t)$ varies significantly for different combinations of $\alpha_H/\alpha{He}$ and $G$, the total SWCX contributions are very similar. We tested different combinations and, as expected, we found that the distribution of R2/R1 ratio after subtracting the SWCX is generally similar, and the systematic uncertainty on the peak of the distribution is $\Delta\frac{R2}{R1}=0.040$, corresponding a temperature difference of $kT$=0.003 keV.
Another possible source of systematics is the fact that the RASS R1 rate is systematically lower than other 1/4 keV all-sky surveys, e.g., the University of Wisconsin sky survey (Dan McCammon, private communication, see also in \citet{McCammon90}). Considering an 18\% correction on the R1 band, to match the RASS results with previous surveys, the peak of the R2/R1 distribution would then be shifted to 0.73, corresponding to a lower temperature of 0.088~keV.

\section{Summary}
Based on the data from the \dxl\ sounding rocket mission, we quantified and removed the SWCX contribution to the foreground diffuse X-ray emission, and obtained a ``cleaned'' map of the LHB emission from the RASS data. Assuming that the LHB is in thermal ionization equilibrium, we measured the temperature of the LHB from the R2/R1 ratio, and estimated its emission measure over the whole sky. We found that the estimated temperature of the LHB is cooler after the contamination of the SWCX is removed. Assuming the LHB has a constant electron density, we also estimated the size of the LHB in each direction and built a three-dimensional model of the LHB, which matches quite well with maps of the local cavity from reddening data.

\begin{acknowledgements}
This work was supported by NASA award numbers
NNX11AF04G and NNX09AF09G.
\end{acknowledgements}

\end{document}